\pdfoutput=1

\documentclass[aps,prl,letterpaper,reprint,nofootinbib,nobibnotes,notitlepage,onecolumn,groupedaddress,superscriptaddress]{revtex4-1}

\usepackage{graphicx}
\usepackage{amsmath}
\usepackage{amssymb}
\usepackage{subfig}
\usepackage{multirow}
\usepackage{hyperref}

\begin{document}


\title{Collective phenomena in granular and atmospheric electrification}


\author{Freja Nordsiek}
\altaffiliation{Now at the Max-Planck Institute for Dynamics and Self-Organization, 37077 G{\"o}ttingen, Germany}
\email{freja.nordsiek@ds.mpg.de}
\affiliation{Departments of Physics and the Institute for Research in Electronics and Applied Physics, University of Maryland College Park, College Park, MD 20742, USA}


\author{Daniel P. Lathrop}
\email[Corresponding author:]{lathrop@umd.edu}
\affiliation{Departments of Physics and the Institute for Research in Electronics and Applied Physics, University of Maryland College Park, College Park, MD 20742, USA}
\affiliation{Joint appointment with the Department of Geology}


\date{\today}

\begin{abstract}
	In clouds of suspended particles (grains, droplets, spheres, crystals, etc.), collisions electrify the particles and the clouds, producing large electric potential differences over large scales.
	This is seen most spectacularly in the atmosphere as lighting in thunderstorms, thundersnow, dust storms, and volcanic ash plumes where multi-million-volt potential differences over scales of kilometers can be produced, but it is a general phenomenon in granular systems as a whole.
	The electrification process is not well understood, especially for electrification of insulating particles of the same material.
	To investigate the relative importances of particle properties (material, size, etc.) and collective phenomena (behaviors of systems at large scales not easily predicted from local dynamics) in granular and atmospheric electrification, we used a table-top experiment that mechanically shakes particles inside a cell where we measure the macroscopic electric field between the electrically conducting end plates.
	The measured electric fields are a result of capacitive coupling and direct charge transfer between the particles and the plates.
	Using a diverse range of mono-material particle sets (plastics, ceramic, glass, and metals), we found that all our particle materials electrify and show similar dynamics with long time-scale temporal variation and an electric field amplitude that depends on the particle quantity in a complex way.
	These results suggest that while particle properties do matter like previous investigations have shown, macroscopic electrification of solids is relatively material agnostic and large scale collective phenomena play a major role.
\end{abstract}


\maketitle

\section{Introduction}

\begin{figure}[b]
	\subfloat[][]{\label{fig:lightningstorm} \includegraphics[width=8cm]{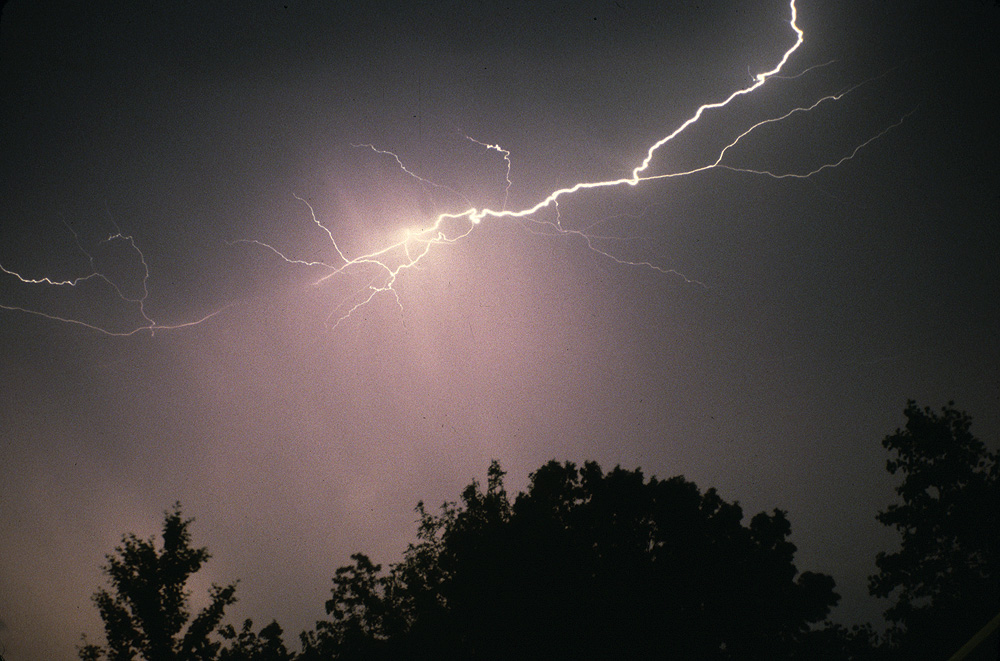}}
	\hspace{1cm}
	\subfloat[][]{\label{fig:volcanic_lightning} \includegraphics[width=7.9cm]{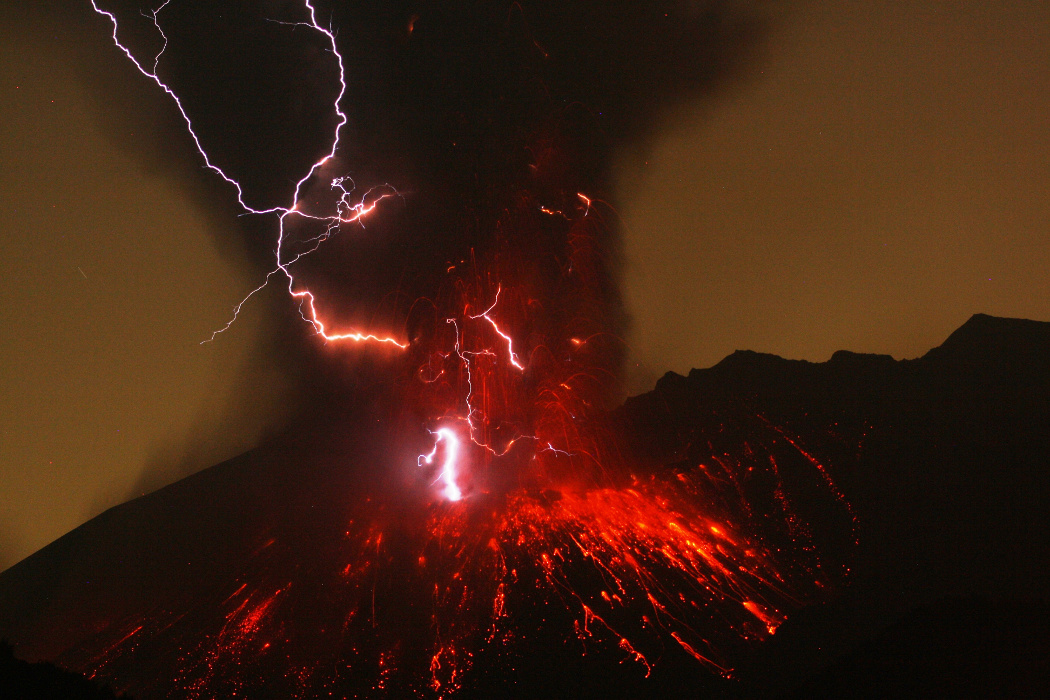}}
	\caption{Natural atmospheric electrification and lightning. \protect\subref{fig:lightningstorm} Lightning in a thunderstorm. Photograph courtesy of John W. Merck, Jr. \protect\subref{fig:volcanic_lightning} Volcanic lightning on Sakurajima. Photograph courtesy of Mike Lyvers.}
	\label{fig:lightning}
\end{figure}

In the atmosphere, clouds of suspended particles electrify \cite{Lacks_Sankaran_JPD_2011,Latham_QuartJournRoyMetSoc_1964,Kok_Renno_PRL_2008,Pahtz_etal_NaturePhysics_2010,Gu_etal_ScientificReports_2013,Cimarelli_etal_Geology_2014,saunders_SpaceScienceReviews_2008,smirnov_PhysicsUspekhi_2014}, leading to the lightning in thunderstorms \cite{Pahtz_etal_NaturePhysics_2010,saunders_SpaceScienceReviews_2008,smirnov_PhysicsUspekhi_2014}, thunder-snow \cite{Latham_QuartJournRoyMetSoc_1964}, dust storms \cite{Lacks_Sankaran_JPD_2011,Latham_QuartJournRoyMetSoc_1964,Kok_Renno_PRL_2008,Pahtz_etal_NaturePhysics_2010,Gu_etal_ScientificReports_2013}, and volcanic ash clouds \cite{Lacks_Sankaran_JPD_2011,Cimarelli_etal_Geology_2014} (Fig.~\ref{fig:lightning}).
This granular electrification process is of interest as a natural phenomenon, but it is also an issue in industry where electrical discharges in flammable powders pose explosive hazards \cite{Lacks_Sankaran_JPD_2011}.
The suspended particles (water droplets, ice, dust, ash, etc.) collide, exchange charge, and transport charge through the system, producing macroscopic electrical potential differences \cite{Lacks_Sankaran_JPD_2011,Latham_QuartJournRoyMetSoc_1964,Kok_Renno_PRL_2008,Pahtz_etal_NaturePhysics_2010,Gu_etal_ScientificReports_2013,Cimarelli_etal_Geology_2014,saunders_SpaceScienceReviews_2008,smirnov_PhysicsUspekhi_2014}.
Past investigations have mainly focused on two-body collisional charge exchange processes.
Surface chemistry, electrical properties, material properties, temperature differences, and surface curvature of and material exchange between the colliding particles have been studied \cite{Lacks_Sankaran_JPD_2011,saunders_SpaceScienceReviews_2008,smirnov_PhysicsUspekhi_2014,Latham_QuartJournRoyMetSoc_1964,Cimarelli_etal_Geology_2014,Lowell_Truscott_JournPhysDAppPhys_1986_1,Lowell_Truscott_JournPhysDAppPhys_1986_2,Duff_Lacks_JournalElectrostatics_2008,Forward_etal_PRL_2009,Forward_etal_GRL_2009,Kok_Lacks_PRE_2009,Angus_etal_JournalElectrostatics_2013,Waitukaitis_etal_PRL_2014,Baytekin_etal_AngewandteChemiInternationalEdition_2012}.
There has also been investigation into the role of background electric fields \cite{Pahtz_etal_NaturePhysics_2010,Siu_etal_PRE_2014}, the presence of other phases of the particle material in the case of H$_2$O \cite{saunders_SpaceScienceReviews_2008,smirnov_PhysicsUspekhi_2014}, and spontaneous charging by local polarization \cite{Yoshimatsu_etal_ScientificReports_2017}.
Despite over a century of observations and investigation of granular electrification \cite{Lacks_Sankaran_JPD_2011,Baddeley_whirlwinds_1860,Rudge_PRSLA_1914}, the actual charging process and the net separation of charges over many kilometers in storms is not well understood in general \cite{Lacks_Sankaran_JPD_2011,Latham_QuartJournRoyMetSoc_1964,Kok_Renno_PRL_2008,Pahtz_etal_NaturePhysics_2010,Gu_etal_ScientificReports_2013,Cimarelli_etal_Geology_2014,Lowell_Truscott_JournPhysDAppPhys_1986_1,Lowell_Truscott_JournPhysDAppPhys_1986_2,Duff_Lacks_JournalElectrostatics_2008,Forward_etal_PRL_2009,Forward_etal_GRL_2009,Kok_Lacks_PRE_2009,Angus_etal_JournalElectrostatics_2013,Waitukaitis_etal_PRL_2014,Siu_etal_PRE_2014,Yoshimatsu_etal_ScientificReports_2017}, though there is better understanding in thunder-snow and thunderstorms cold enough to have ice where all three phases of H$_2$O are present \cite{saunders_SpaceScienceReviews_2008,smirnov_PhysicsUspekhi_2014}.
A particularly difficult case is the electrification of insulating particles \cite{Lacks_Sankaran_JPD_2011,Pahtz_etal_NaturePhysics_2010, Lowell_Truscott_JournPhysDAppPhys_1986_1,Lowell_Truscott_JournPhysDAppPhys_1986_2,Duff_Lacks_JournalElectrostatics_2008,Forward_etal_PRL_2009,Forward_etal_GRL_2009,Kok_Lacks_PRE_2009,Angus_etal_JournalElectrostatics_2013,Waitukaitis_etal_PRL_2014,Siu_etal_PRE_2014,Yoshimatsu_etal_ScientificReports_2017}, which are found in the industrial setting and in dust storms.

\textit{
In this Letter, we describe a table-top experiment designed to study the collective phenomena that underlie granular electrification.
We then describe initial data that validate the use of this apparatus for such studies, and discuss what the initial data with mono-material particle sets for a diverse range of materials (plastics, ceramic, glass, and metals) implies for granular electrification in general and in the atmosphere.
Specifically, we found electrification in all particle types we tested and all particle types showed similar dynamics that vary on long time-scales with voltage amplitudes that depend on the number of particles in a complex way.
This suggests collective phenomena play a key role in granular electrification and that material properties are less important than currently thought.
}

Collective phenomena are behaviors of physical systems at large scales, not easily predicted by the local dynamics.
Equilibrium examples include the phases of matter, thermodynamic phase transitions, and critical points.
Non-equilibrium examples include fluid turbulence, jamming in granular flows, swarming behavior in animals, and pattern-forming systems.
In granular electrification, there are several potential sources for collective phenomena.
For example, the macroscopic rearrangement of particles with electrical charge is a prerequisite for electric potential differences to become large enough for a discharge.
As well, there should be effects on two-particle contact charging due to the macroscopic electric field of the collection of particles \cite{Pahtz_etal_NaturePhysics_2010,Siu_etal_PRE_2014,Yoshimatsu_etal_ScientificReports_2017}.
We take the first step toward elucidating and delineating collective effects in particle electrification by arguing \textit{they are needed to explain both laboratory and natural phenomena}.

Data has been deposited at \citet{nordsiek_lathrop_SingleParticleType_DRUM_2015}.

\section{Experiment}

\subsection{Apparatus}

\begin{figure}[b]
		\subfloat[][]{\label{fig:experiment_diagram} \includegraphics[width=5.5cm]{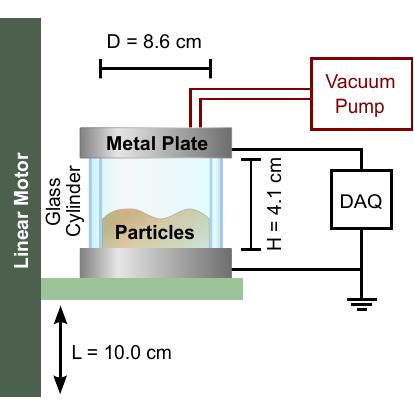}}
		\hspace{1cm}
		\subfloat[][]{\label{fig:experiment_picture} \includegraphics[width=8cm]{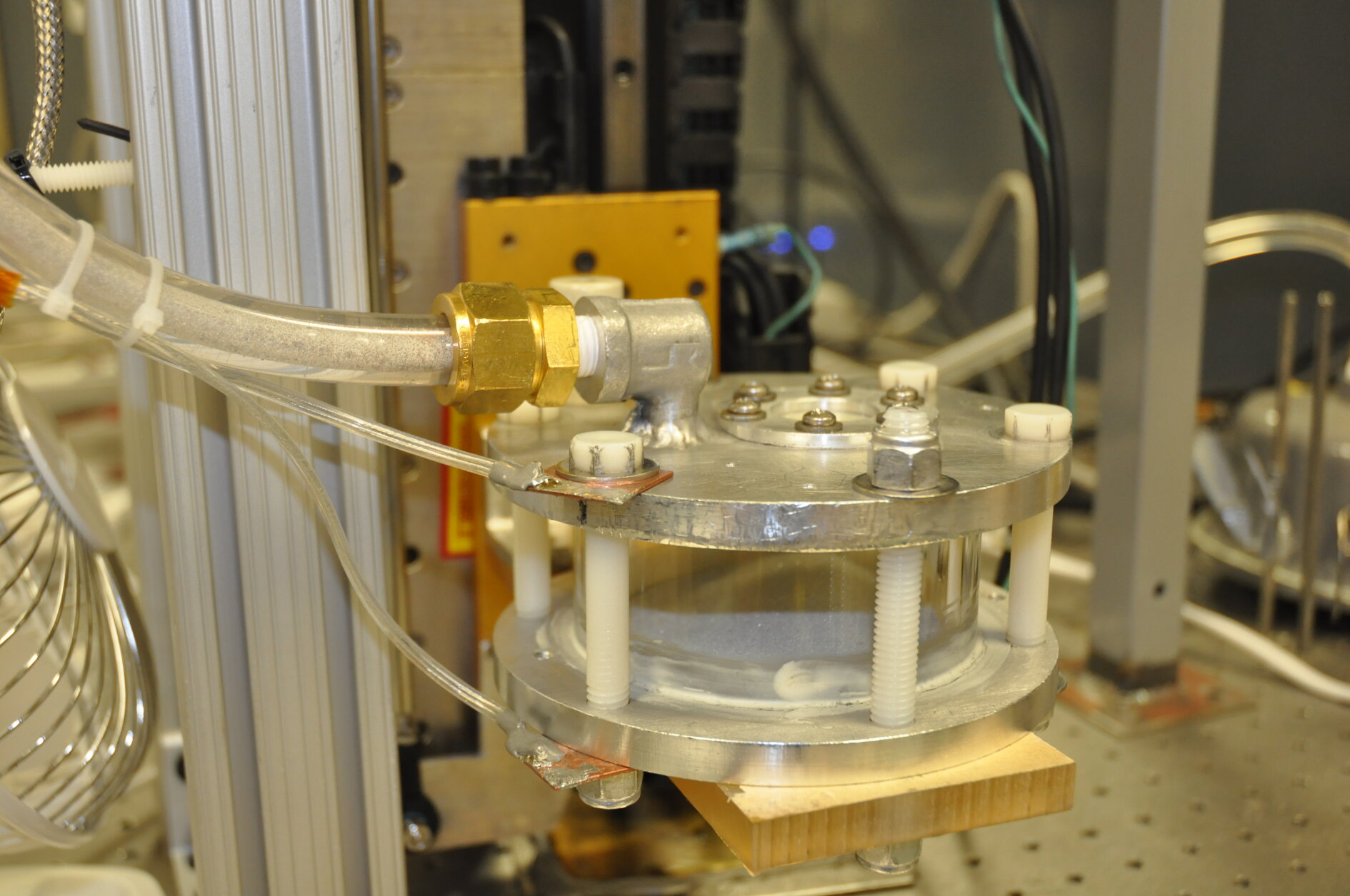}}
	\caption{Experimental \protect\subref{fig:experiment_diagram} diagram and \protect\subref{fig:experiment_picture} photograph. The experiment consists of the cell containing granular particles, the linear servo-motor that shakes the cell vertically, a vacuum pump, and the data acquisition system (DAQ).}
	\label{fig:experiment}
\end{figure}

We investigated granular electrification in a table-top shaking experiment, shown in Fig.~\ref{fig:experiment}.
Thousands to millions of particles are shaken vertically inside a cylindrical cell with conductive top and bottom plates and an insulating sidewall.
The experimental system we present in this paper is an upgrade from the previous work by Paul Lathrop, Daniel Lathrop, Zack Lasner, Julia Salevan \citep{Salevan_thesis_2012}, Tyler Holland-Ashford, and Allison Bradford.
The upgrades consisted of replacing the vacuum system, redoing the electronics and acquisition, redoing the control and acquisition software, and a few small changes to the cell itself.
Additional details are described in the PhD thesis of \citet{Nordsiek_PhD_thesis_2015}.

The cylindrical sidewall is a borosilicate glass (pyrex) cylinder (G. Finkenbeiner) with a $10.0$~cm outer diameter, $D = 8.6$~cm inner diameter, and $5.0$~cm height.
The end caps are two $13.67$~cm diameter and $1.0$~cm thick circular aluminum plates (top and bottom) with annular slots to accommodate the glass sidewall and seal the cell.
Six insulating nylon bolts hold the plates tightly together with the glass sidewall bound between them.
Due to the annular slots in the plates, the interior of cell, where the particles shake, has a diameter of $D = 8.6$~cm and a height of $H = 4.12$~cm.

Humidity affects electrification, as is seen by the greater ease of generating static electricity in arid climates and the dry of winter.
To reduce this effect, the cell is held under a partial vacuum of $30$--$60$ mTorr ($4 \times 10^{-5}$ to $8 \times 10^{-5}$~atm).
For comparison, the saturated vapor pressure of water at the approximate lab temperature of $23$~C is $21.08$~Torr \cite{lide_crc_2000}.
This does, however, make our experiment deviate from atmospheric electrification since the impact of the remaining gas on the particle trajectories in our experiment is negligible, unlike the atmospheric case.
The cell is continuously pumped through a tube fitting and hole in the top plate.
The hole is covered in a metal mesh to keep particles from getting into the vacuum system.
The pressure was measured behind the last filter by a convection vacuum gauge (Duniway CVT-275-101 with a Granville-Phillips 275 Analog Convectron Gauge Controller).
The inside of the top plate is not flat due to the mesh, a sheet of aluminum covering a previously cut window, and seven small round-head bolts holding the former two items in place.

The cell is shaken vertically by a linear motor with a stroke-length of $L = 10.0$~cm using a square-wave acceleration profile with amplitude $a$, which causes the cell's vertical position to be a sequence of connected parabolas of opposite concavity.
This was done by attaching the cell to a linear servo-motor (Trilogy T2SA19-3NCTS motor controlled by a Parker Hannifin Gemini GV6-U3E-DN servo-motor drive) with an electrically insulating plate of acrylic.
The linear servo-motor was mounted so that motion is vertical, with a maximum stroke-length of $24.13$~cm.
The vertical position of the cell was measured by a continuous linear potentiometer (Penny+Giles SLS190/0300/L/66/01/N) with a $30$~cm stroke.
As the cell could implode due to the partial vacuum it contains, the experiment is enclosed in an acrylic shield.
A small desk fan was placed inside the shield to keep the motor cool when it is run for long periods of time.

\subsection{Particles}

\begin{table}[b]
	\caption{Detailed information on the particles. Composition, manufacturer, and size information for all particle types used. The spheres come with a manufacturer nominal diameter range. The powders were characterized and the statistics of their effective diameters shown here. RMS stands for Root Mean Squared, and STD stands for Standard Deviation. We show the excess kurtosis.}
	\begin{ruledtabular}
	\begin{tabular}{llllllllllll}
		& & & & & \multicolumn{7}{c}{Effective diameter statistics ($\mu$m)} \\
		\cline{6-12}
		Material &	Form &	Manufacturer &	Product \# &	Nominal diameter ($\mu$m) &	Min. &	Max. &	Mean &	RMS &	STD &	Skew &	Kurtosis \\
		\hline
		Aluminum &	Powder &	GoodFellow &	AL006010 &	Max. 400 &	58 &	376 &	231 &	242 &	74 &	-0.308 &	-0.408 \\
		Copper &	Powder &	GoodFellow &	CU006045 &	Max. 200 &	5 &	166 &	43 &	49 &	24 &	1.09 &	1.84 \\
		Glass &	Spheres &	GlenMills &	7200-000200 &	200--300 &	- &	- &	- &	- &	- &	- &	- \\
		Glass &	Spheres &	GlenMills &	7200-000400 &	400--600 &	- &	- &	- &	- &	- &	- &	- \\
		Glass &	Spheres &	GlenMills &	7200-000750 &	750--1000 &	- &	- &	- &	- &	- &	- &	- \\
		Polystyrene &	Powder &	GoodFellow &	ST316051 &	Max. 250 &	12 &	324\footnote{There was one outlier, which was excluded, whose effective diameter was 438 $\mu$m.} &	111 &	131 &	70 &	1.2 &	1.75 \\
		Polystyrene &	Spheres &	GlenMills &	7192-PB-2 &	360--610 &	- &	- &	- &	- &	- &	- &	- \\
		Polystyrene &	Spheres &	GlenMills &	7192-PB-1 &	610--990 &	- &	- &	- &	- &	- &	- &	- \\
		PTFE &	Powder &	GoodFellow &	FP306068 &	Max. 675 &	35 &	636\footnote{There was one outlier, which was excluded, whose effective diameter was 1037 $\mu$m.} &	212 &	303 &	218 &	1.88 &	3.84 \\
		ZrO$_2$:SiO$_2$ &	Spheres &	GlenMills &	7305-000002 &	200--300 &	- &	- &	- &	- &	- &	- &	- \\
		ZrO$_2$:SiO$_2$ &	Spheres &	GlenMills &	7305-000004 &	400--600 &	- &	- &	- &	- &	- &	- &	- \\
		ZrO$_2$:SiO$_2$ &	Spheres &	GlenMills &	7305-000010 &	800--1000 &	- &	- &	- &	- &	- &	- &	- \\
	\end{tabular}
	\end{ruledtabular}
	\label{table:particles}
\end{table}

We used 12 types granular particles of different materials, forms (spheres and powders), and diameter ranges with the largest diameter being under $1$~mm.
The particle types were aluminum powder, copper powder, lead-free soda-lime glass spheres, polystyrene powder and spheres, PTFE (polytetrafluoroethylene) powder, and 69\%:31\% ZrO$_2$:SiO$_2$ spheres.
These particle types include a total of two plastics, one glass, one ceramic, and two metals.
All twelve types of particles we used are summarized in Table~\ref{table:particles}.

\begin{figure}
	\subfloat[][]{\label{fig:al} \includegraphics[width=4cm]{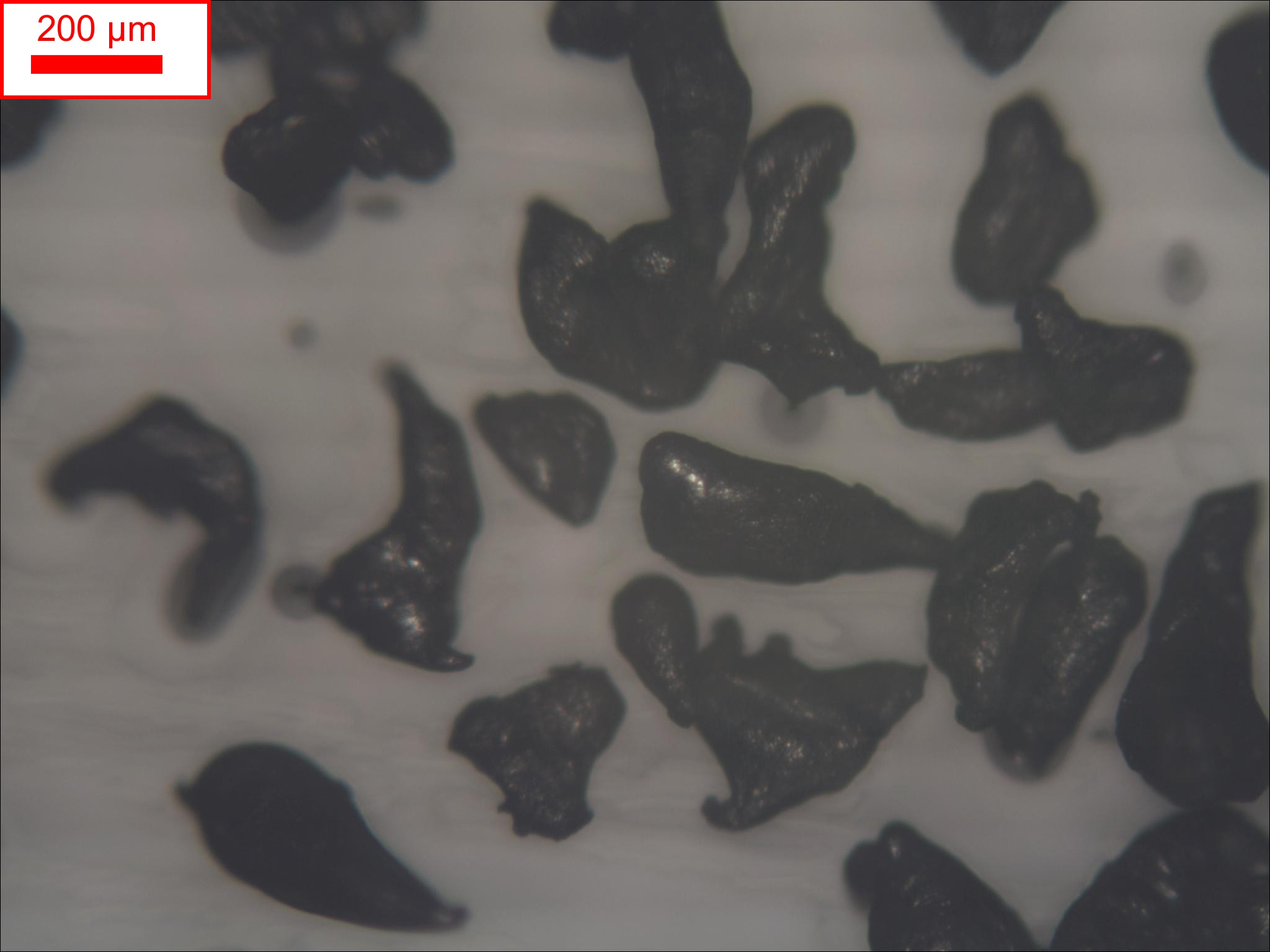}}
	\subfloat[][]{\label{fig:cu} \includegraphics[width=4cm]{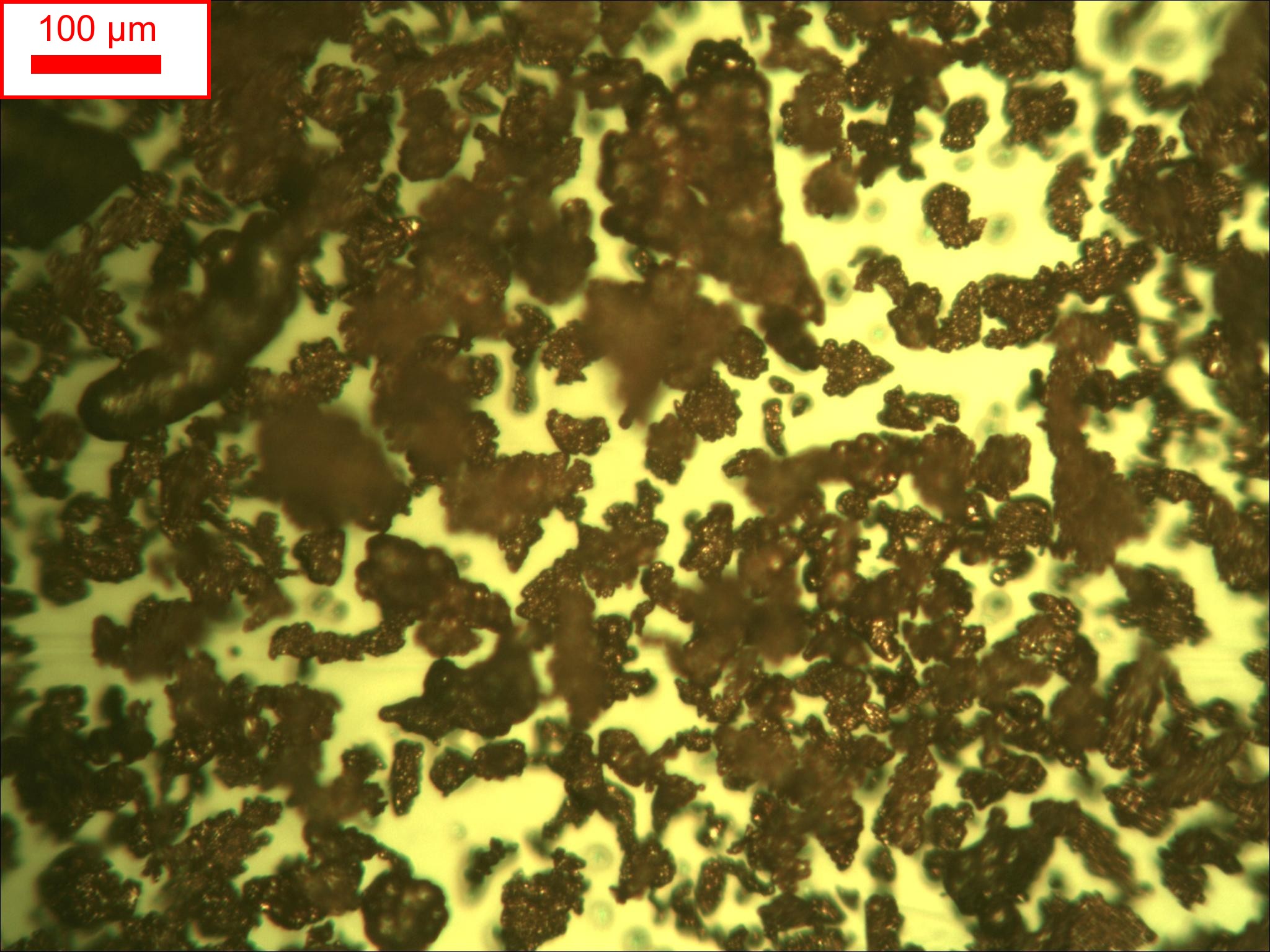}}
	\subfloat[][]{\label{fig:ps} \includegraphics[width=4cm]{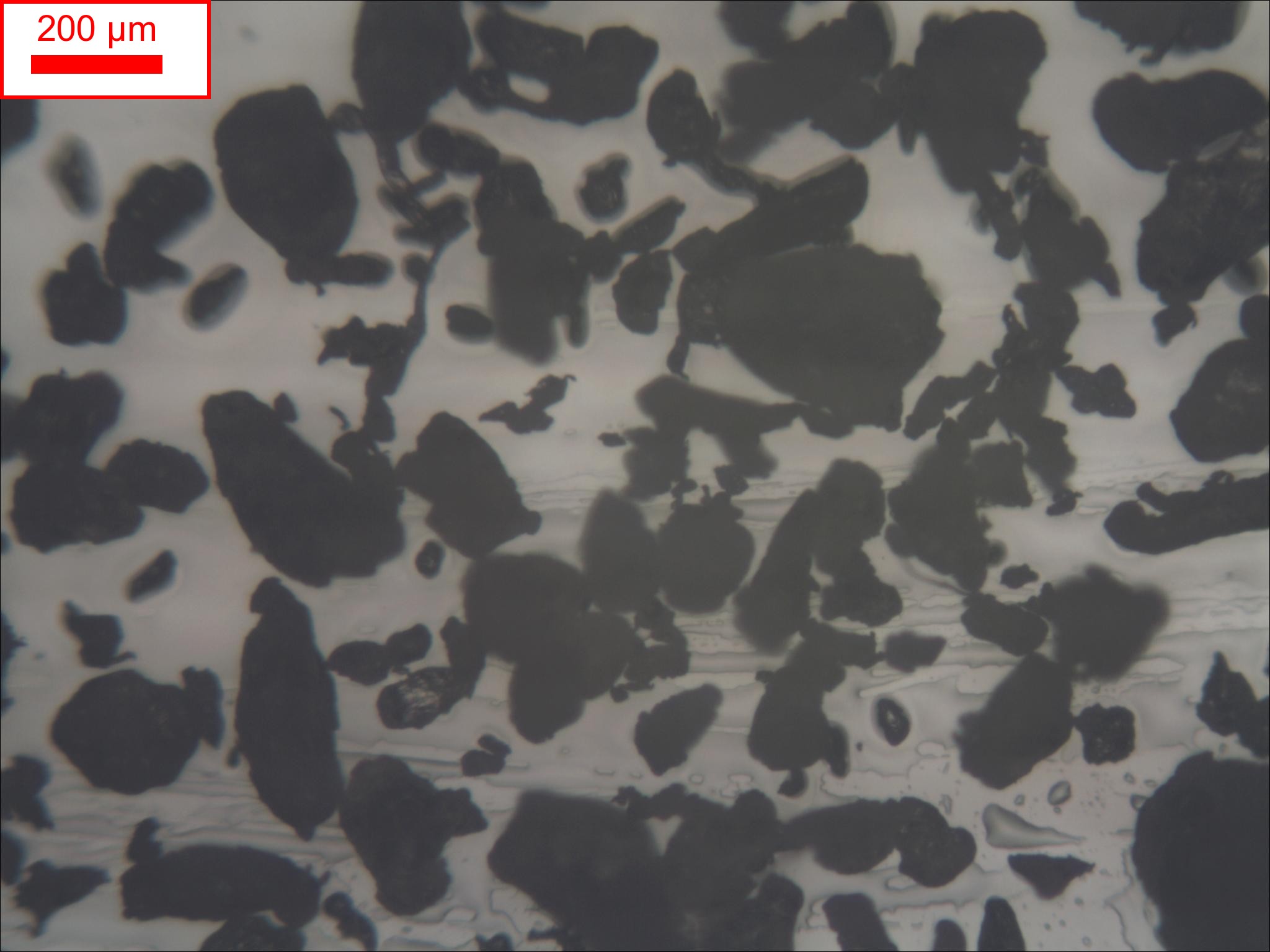}}
	\subfloat[][]{\label{fig:ptfe} \includegraphics[width=4cm]{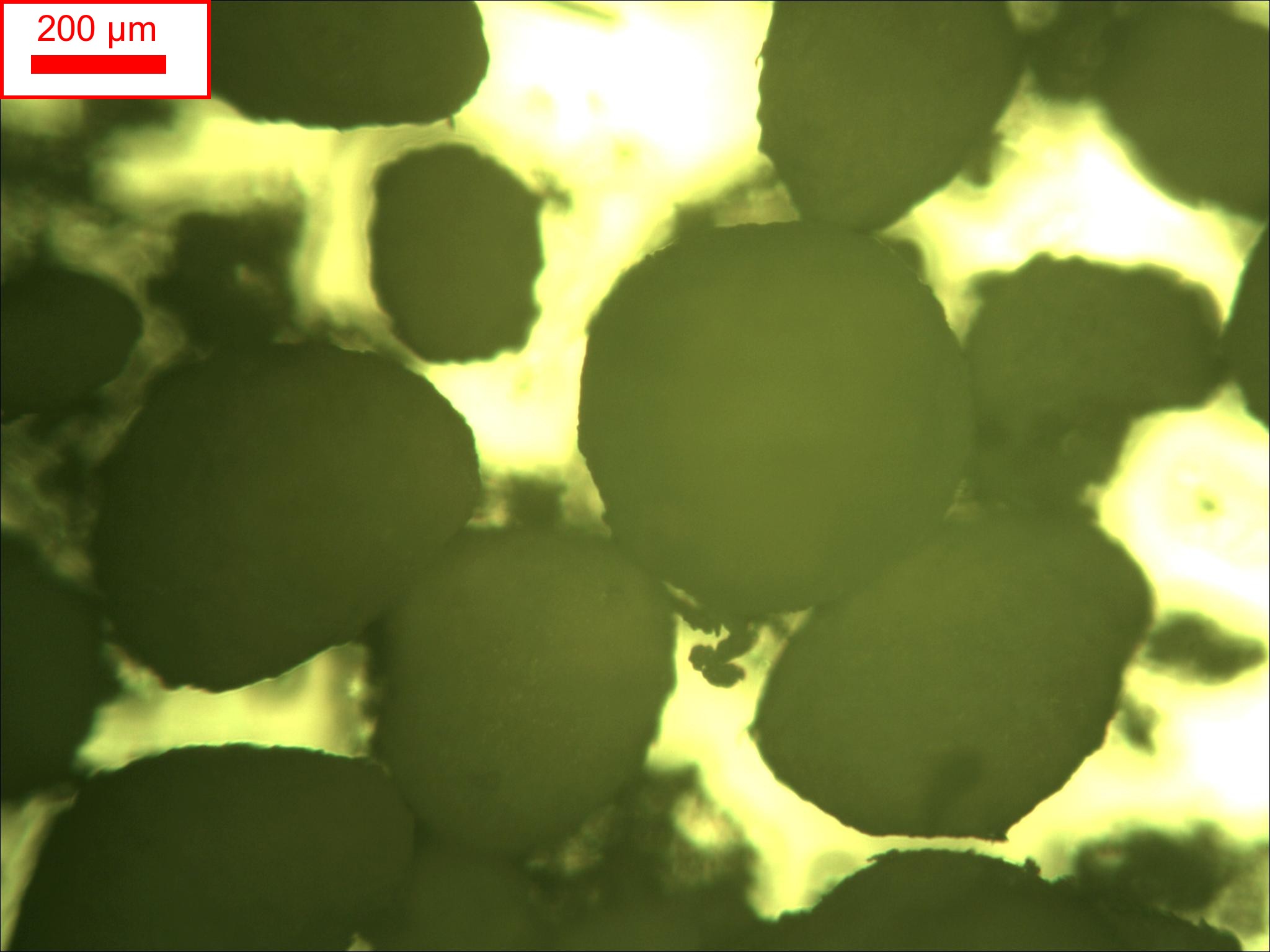}}
	\\
	\subfloat[][]{\label{fig:h_al} \includegraphics{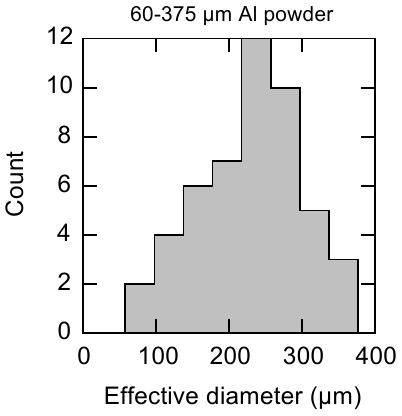}}
	\subfloat[][]{\label{fig:h_ptfe} \includegraphics{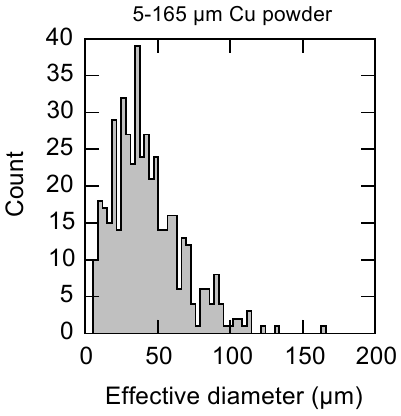}}
	\subfloat[][]{\label{fig:h_cu} \includegraphics{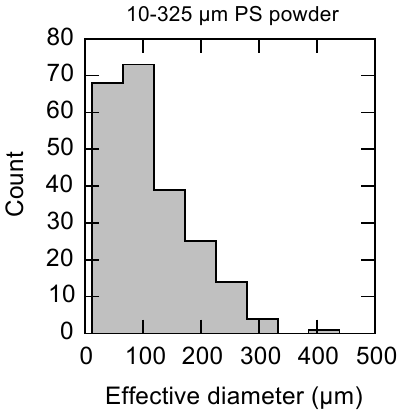}}
	\subfloat[][]{\label{fig:h_ps} \includegraphics{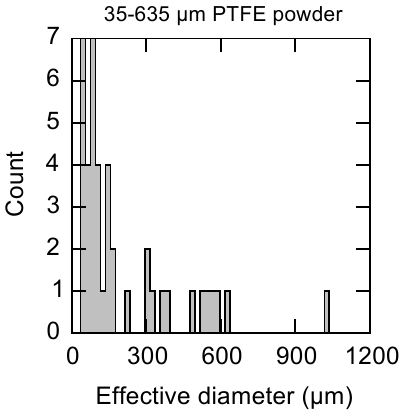}}
	\caption{Powder microscopy and effective diameter distributions. One microscope (Leitz Ergolux) image (courtesy of John Abrahams III) of each of the powders used in this paper and histograms of their effective diameters (Materials and Methods). \protect\subref{fig:al} and \protect\subref{fig:h_al} 60--375 $\mu$m aluminum, which had 49 particles characterized. \protect\subref{fig:cu} and \protect\subref{fig:h_cu} 5--165 $\mu$m copper, which had 457 particles characterized. \protect\subref{fig:ps} and \protect\subref{fig:h_ps} 10--325 $\mu$m polystyrene (PS), which had 224 particles characterized. \protect\subref{fig:ptfe} and \protect\subref{fig:h_ptfe} 35--635 $\mu$m PTFE, which had 42 particles characterized.}
	\label{fig:powders}
\end{figure}

Each powder was characterized by taking three images under a microscope (Leitz Ergolux) while backlit.
The cross-sectional areas of the particles were obtained by outlining each particle by hand with interpolation for overlapping particles in an image editor and counting the number of pixels.
The diameter of a circle with the same cross-sectional area was defined to be the effective diameter of the particle.
One image of each powder and histograms of their effective diameters are shown in Fig.~\ref{fig:powders}.

The particles, whether spheres or powders, were poly-disperse.
For the spheres, the range of particle diameters was 20-50\% of the maximum diameter.
For the powders, the effective diameters of the largest particles are an order of magnitude larger than for the smallest particles.

There are several different ways to quantify the amount of granular material in the cell.
We decided to use the thickness of the slab of particles that forms at the bottom of the cell when the cell is at rest, and make it dimensionless.
Ignoring the voids between the particles, the thickness of the slab at rest, $h$, is the total particle volume $\tfrac{1}{6} \pi N_p d^3$ divided by the cross-sectional area of the cell $\tfrac{1}{4} \pi D^2$, which is

\begin{equation}
	h = \frac{2 N_p d^3}{3 D^2} \quad ,
	\label{eqn:slab_thickness}
\end{equation}

\noindent where $N_p$ is the number of particles, $d$ is the particle diameter, and $D$ is the diameter of the cell.
We then define the dimentionless thickness of the slab to be

\begin{equation}
	\lambda = \frac{h}{d} = \frac{2 N_p d^2}{3 D^2} \quad .
	\label{eqn:lambda}
\end{equation}

\noindent This is proportional to the total surface area of the particles, which is $4 \pi N_p d^2$.
Considering that electrification is a process that happens on the particle surfaces, this is a useful property to have.
In addition, $\lambda$ is approximately how many mono-layers of particles are in the cell since it is the height of the slab in units of $d$.

The mass of particles is measured on a scale and $\lambda$ is calculated using their diameter and densities, or the target mass is calculated for a desired $\lambda$ and then that quantity is measured out.
We use the manufacturer provided particle densities.
For $d$, we use the mean of the manufacturer provided diameter bounds for spherical particles and the RMS (Root Mean Squared) effective diameter for powders.

\subsection{Electronics and Data Acquisition}

We acquired the electric potential between the plates and the vertical position of the plate.

The top and bottom plates are connected by separate high voltage wires (insulation rated to $12$~kV) to the acquisition system and ground respectively.
Since potentials up to $1$~kV from electric discharges inside the cell had been previously measured across the cell before the work presented in this paper, the electric potential of the top plate was stepped down by a factor of $11.0$ for normal operation and $101.0$ for measuring discharges.
They were stepped down using two high impedance resistors of $10.0$~M$\Omega$  and $1.00$~M$\Omega$ ($100.0$~k$\Omega$ for discharge measurements) before passing through an instrumentation amplifier (Analog Devices AD624).
With regulated DC rails of $\pm 9$~V, measurements are clipped to $\pm 70$~V ($\pm 600$~V for discharge measurements).
The instrumentation amplifier provides a $20$~pF capacitance between the input for the top plate and ground.
An additional $10$~pF capacitor is connected across the $10.0$~M$\Omega$ resistor.
With a plate capacitance of $3.15$~pF, the total capacitance between the top plate and ground is $13$~pF if we approximate the connection to the instrumentation amplifier as ground since the drain resistor between it and ground is $11.0$ or $101$ times smaller.
The RC time constant for the potential difference is then $\approx 130$~$\mu$s.

The vertical position sensor (linear potentiometer) was operated as a voltage divider with a fixed regulated $5$~V across it and the potential from the center tap was measured by the acquisition system through an instrumentation amplifier.
To calibrate the vertical position measurement with the linear potentiometer, the cell was set to 21 vertical positions over the full range and the potentials across the whole linear potentiometer and at the center tap were acquired at each step.
The linear servo-motor itself has a precise position encoder which allows the cell's position to be set precisely programmatically, though it cannot be read in real time while the motor is in motion.
A linear fit is done between the measured potentials at the center tap and the positions of each step.
The potential from the regulated power supply across the whole potentiometer is approximately constant.

The outputs of the two instrumentation amplifiers are read by a high speed DAQ (Data AcQuisition system) at 80 kHz each (National Instruments USB-6211).
To reduce interference and noise, standard practices were used such as twisted shielded wire, an electrically conductive enclosure for the electronics, no ground loops in the acquisition electronics, regulated AC-DC power, and isolation between the acquisition computer, motor drive, and the acquisition electronics.

\section{Results}

\subsection{Basic Electrification}

%

\begin{figure}
	\begin{tabular}{ccc}
		\setcounter{subfigure}{1}
		& \multirow{-2}[35]{*}{\subfloat[][]{\label{fig:discharge} \includegraphics{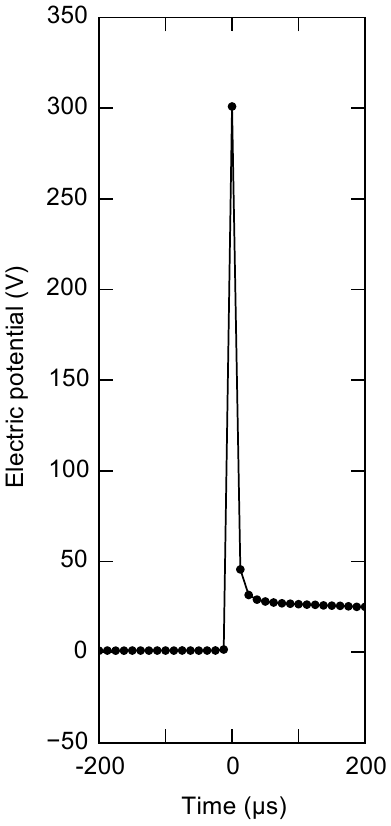}}} & \subfloat[][]{\label{fig:timetrace} \includegraphics{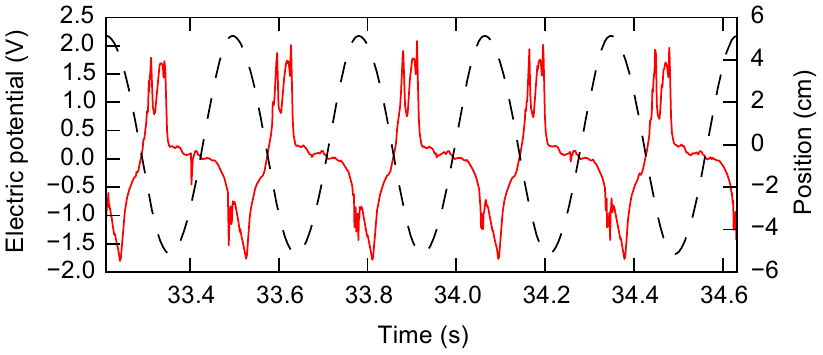}} \\
		\setcounter{subfigure}{0} \subfloat[][]{\label{fig:electrification_methods} \includegraphics{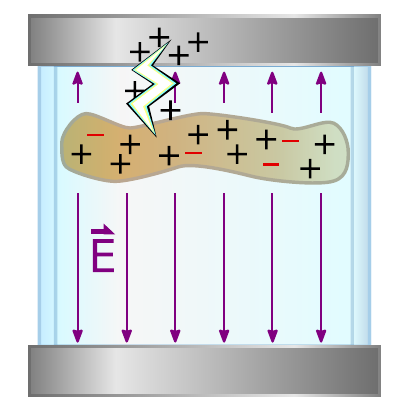}} &  & \setcounter{subfigure}{3} \subfloat[][]{\label{fig:std_cycles} \includegraphics{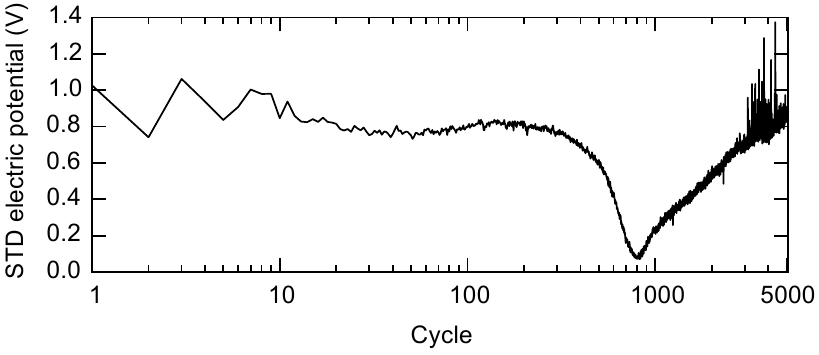}}
	\end{tabular}
	\caption{\protect\subref{fig:electrification_methods} A schematic of how an electric potential difference between the plates is established. Capacitive coupling from the charges in the slab of particles (light brown) cause electric fields (purple lines) and potential differences between the plates. There is also direct transfer of charge to the plates by electric discharge (drawn as a lightning bolt). \protect\subref{fig:discharge} Time trace of the electric potential across the cell during an electrical discharge. \protect\subref{fig:timetrace} \protect\subref{fig:std_cycles} Electric potential difference time dependence for 400--600 $\mu$m glass spheres. Time dependence for a quantity  $\lambda = 6$ (Eq.~\ref{eqn:lambda}) of 400--600 $\mu$m diameter glass spheres shaken at $a = 2.08 \, g$ for 5000 cycles. \protect\subref{fig:timetrace}  The cell vertical position (black dashed line) and electric potential (red solid line) time series for cycles 100--105. \protect\subref{fig:std_cycles}  Standard deviation of the electric potential for each cycle. There is an inversion around cycle 800.}
	\label{fig:electrical_and_examples}
\end{figure}

We quantified the electrification by the electric potential between the plates (bottom plate is defined as $0$~V).
While the cell is shaking, the particles form a loose slab that collapses inelastically when it hits either plate.
The particles can collisionally exchange charge with each other, the metal plates, and the glass sidewall.

There are two primary mechanisms that can create an electric potential difference between the plates, which are illustrated in Figure~\ref{fig:electrification_methods}.
First, there can be capacitive coupling between particle charges and the plates while the particles are in flight.
Second, the particles can electrically discharge to a plate, depositing charge directly onto it, which may cause much larger potential differences.
When the cell is shaken, we observe an oscillating electric potential across the cell  (Figure~\ref{fig:timetrace}), typically a few volts peak to peak.
There are also occasional multi-hundred volt discharges such as in Figure~\ref{fig:discharge}, which can be of either sign.
From Figure~\ref{fig:timetrace}, the electric potential typically has voltage extrema with opposite signs correlated with the extrema in the cell's position (top and bottom of its vertical motion).

While the electric potential is not quite symmetric on both sides of 0~V, we can still describe it as either in-phase with the position (maxima near the top of the cell's motion and a minima near the bottom) or out-of-phase with the position.
We call changes between in-phase and out-of-phase \textit{inversions}.
After the particles have rested for long periods of time, there is an initial transient of a few tens of cycles when starting to shake.
Then, from one cycle to the next, the potential's time dependence changes little (Figure~\ref{fig:timetrace}).
However, over longer time scales the variation is significant (Figure~\ref{fig:std_cycles}), including inversions.
For example, in Figure~\ref{fig:std_cycles}, there is an inversion around cycle $800$ where the standard deviation of the electric potential approaches zero.

\subsection{Cycle Profiles of The Potential}

We will now break our time series into the individual shaking cycles to get the cycle profiles.
Using the tops of the oscillation cycles as the cycle boundaries, all the acquired data points between successive tops are averaged into $1$~ms bins.
Since the cell starts and ends at the bottom of the shaking cycle, half a cycle is discarded from the beginning and end.
For shaking with a square-wave acceleration profile of magnitude $a = 2.08 \, g$ (shaking frequency of $3.522$~Hz) and stroke-length $10.0$~cm, this corresponds to 284 bins in each cycle (about 80 samples per bin at the $80$~kHz acquisition rate).
The tops and bottoms of the oscillation cycles (extrema in the vertical position signal) were found by taking the full vertical position time series before it is put into 1~ms long bins, subtracting the mean, finding all intervals in the time series where the position is more than 1~cm away from the mean position continuously for at least 10~ms, and the extremum taken to be the location of the maximum/minimum position value in the interval.

The measured vertical cell position profile gives an acceleration profile that is approximately a square-wave.
We get the acceleration amplitude, $a$, from $a T^2 = 4 L$ using the measured peak-peak vertical position $L$ and oscillation period $T$.

It is helpful to consider the particles in their flight between the plates using a simple kinematic model.
Using the motion and geometry of the cell, we simulate the ballistic motion of a hypothetical inelastic point particle within the cell.
Being inelastic means that it does not bounce from the plates nor does it have any adhesive forces between it and the plates.
Let $z_c$ and $z_p$ be the vertical positions of the cell bottom and the particle respectively.
We define $z_c(t)$ for the profile over a cycle of the measured vertical position.
We simulate the hypothetical particle's motion using its equations of motion.
Its equations of motion when against the top and bottom plates are, respectively,

\begin{equation}
	\begin{array}{r c l l l}
		z_p & = & z_c + H & \quad \text{when $z_p = z_c + H$ and $\ddot{z}_c \le -g$} \\
		z_p & = & z_c & \quad \text{when $z_p = z_c$ and $\ddot{z}_c \ge -g$}
	\end{array}
\end{equation}

\noindent Otherwise, the particle is in free fall between the plates moving as

\begin{equation}
	\ddot{z}_p = -g \quad .
\end{equation}

\noindent From the simulated trajectories, we find when the particle hits and leaves the two plates, which are the beginnings and ends of periods of time when $z_p = z_c, \, z_c + H$.

\begin{figure}
	\centering{\includegraphics{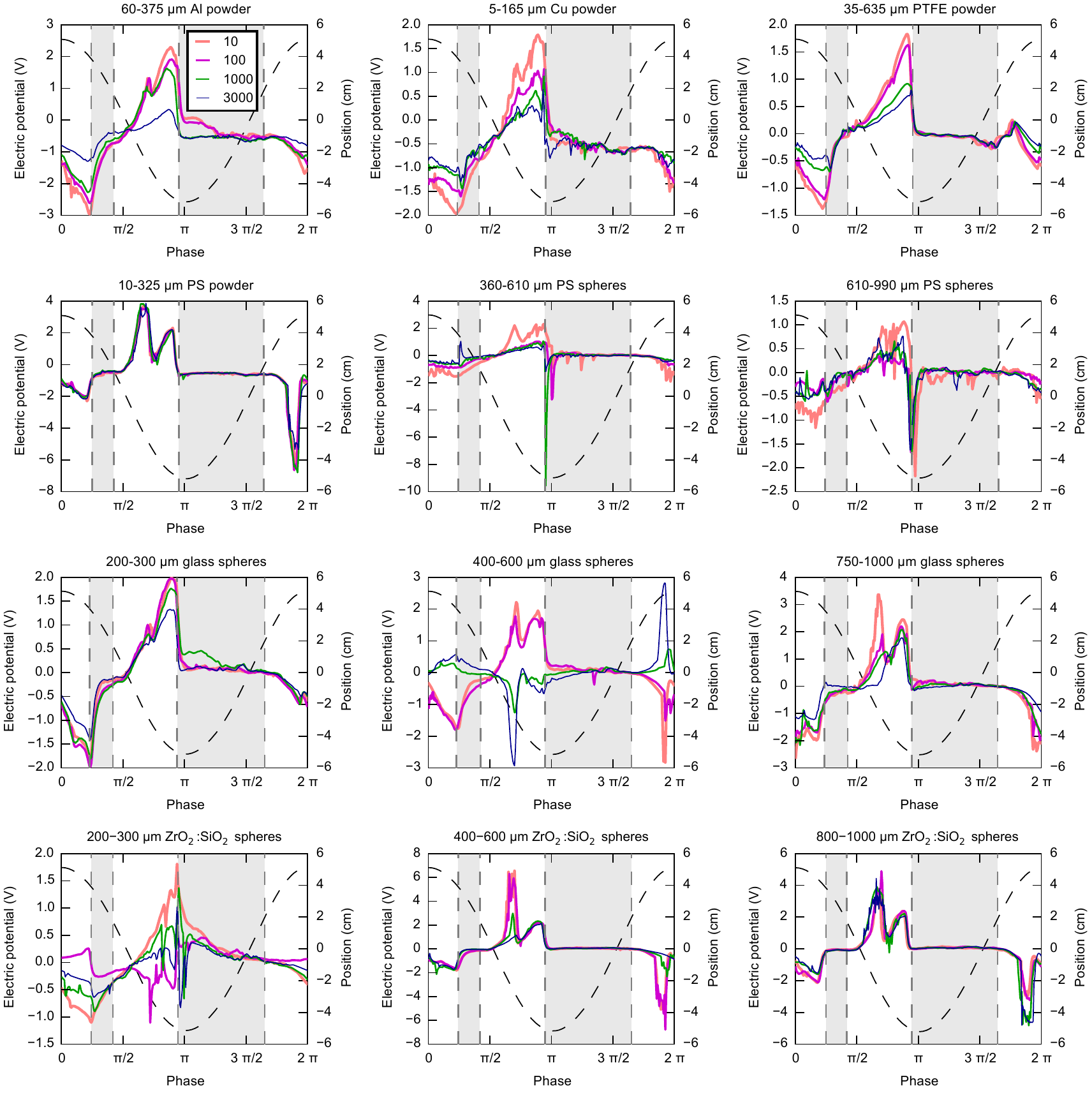}}
	\caption{Signal profiles during the shaking cycle by particle type. The cell vertical position (black dashed line) and electric potential for cycles 10, 100, 1000, and 3000 (legend in \textit{Top-Left} figure) are shown by phase in the shaking cycle for $\lambda = 6$ of each particle type (figure titles) shaken 5000 cycles with $a = 2.08 \, g$. A particle in the kinematic model is on either plate in the grey regions, and the dashed vertical lines, from left to right, are when it hits the top plate, leaves the top plate, hits the bottom plate, and leaves the bottom plate respectively. Polystyrene is abbreviated as PS.}
	\label{fig:cycle_profiles}
\end{figure}

For all 12 particle types, Fig.~\ref{fig:cycle_profiles} shows cycle profiles of the potential across the cell for cycles $10$, $100$, $1000$, and $3000$ shaken at $a = 2.08 \, g$ (shaking frequency of $3.522$~Hz).
The cell's vertical position is also shown.
We mark where this hypothetical particle hits and leaves each plate using dashed vertical lines, and the periods of time where the hypothetical particle is in contact with either plate as the grey shaded regions.

The profile shapes for all particle types are qualitatively similar.
Some patterns are:
\begin{enumerate}
	\item The amplitudes of the oscillating potential are within an order of magnitude of each other for all particle types.
	\item The electric potential trends towards a DC value close to zero while the particles are in contact with either plate.
	\item The profiles show an extremum in the potential right before hitting the top plate.
	\item The profiles show another extremum of the opposite sign before particles hit the bottom plate, with a decay towards DC value close to zero potential afterwards that is usually quicker.
	\item There is often an extremum when the particles are in mid-flight after leaving the top plate before they hit the bottom plate.
	\item Less often, there is an extremum when the particles are in mid-flight before they reach the top plate after leaving the bottom plate.
\end{enumerate}

\subsection{Particle Amount}

We find a strong dependence on the observed electrification due to particle quantity.
To investigate the $\lambda$ dependence of electrification, we shook polystyrene powder samples with different $\lambda$ at different accelerations for $1000$ or $500$ cycles.
And we did the same for polystyrene spheres and glass spheres, but at only a single acceleration $a = 2.08 \, g$ (shaking frequency of $3.522$~Hz).
To improve consistency and reduce the effects of initial transients when investigating the $\lambda$ dependence of electrification, we shook each particle sample first for $10,000$ cycles at  $a = 2.08 \, g$.
Then, we shook the particles at the desired $a$ for $1000$ or $500$ cycles, which was taken to be our results.
If the $\lambda$ dependence was measured for a particle sample at more than one $a$, we shook the particles for at least $100$ cycles at $a = 2.08 \, g$ between each measurement.

\begin{figure}
	\subfloat[][]{\label{fig:lambda_dependence_ps_powder} \includegraphics{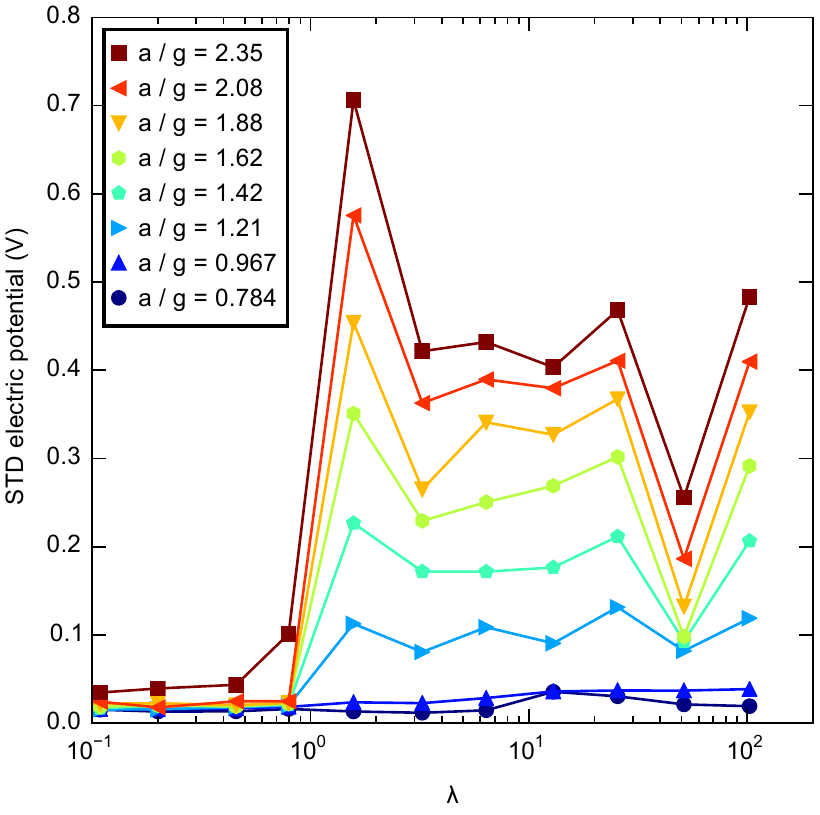}}
	\subfloat[][]{\label{fig:lambda_dependence_three_materials} \includegraphics{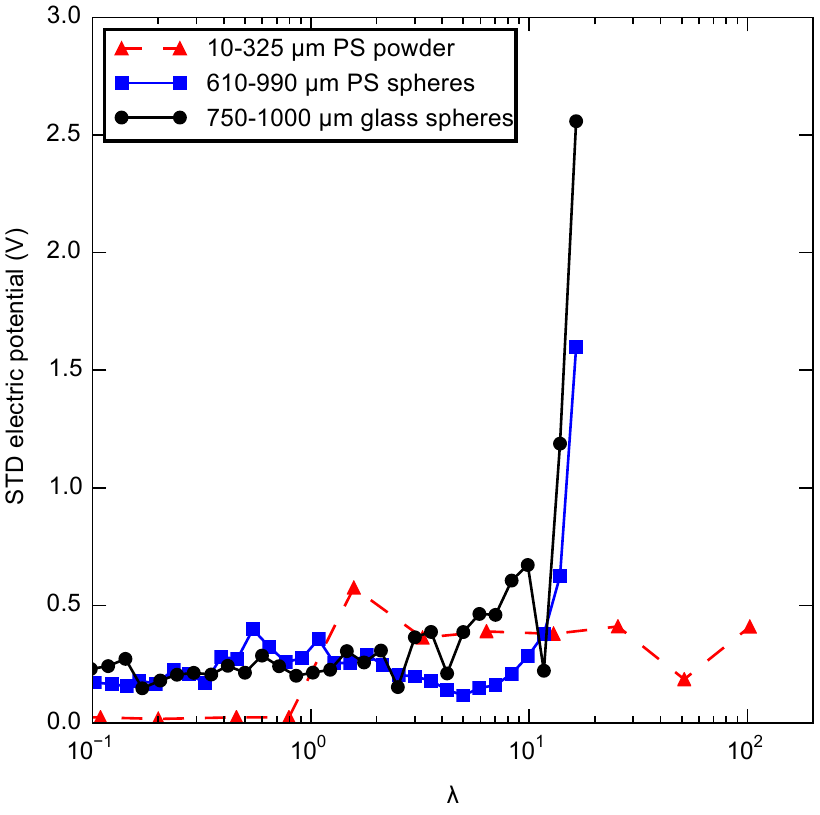}}
	\caption{Strength of electric potential by particle quantity and acceleration strength for different particle types. We show standard deviation of the measured potential between the plates for different quantities, $\lambda$ and $a / g$, shaken for 1000 or 500 cycles (the first 10 cycles were skipped to remove the initial transient). \protect\subref{fig:lambda_dependence_ps_powder} Different acceleration strengths for 10--325~$\mu$m polystyrene powder. \protect\subref{fig:lambda_dependence_three_materials} Three different particle types (polystyrene abbreviated as PS) shaken at $a = 2.08 \, g$ (shaking frequency of $3.522$~Hz.}
	\label{fig:lambda_dependence}
\end{figure}

The standard deviations of the electric potentials between the plates (skipping the first $10$ cycles) are shown in Fig.~\ref{fig:lambda_dependence}.
First, we consider the polystyrene powder (Fig.~\ref{fig:lambda_dependence_ps_powder}).
The number of particle-particle collisions increases with increasing $\lambda$ and $a / g$ (forcing strength), and they become more energetic for increasing $a / g$.
The measured potential increases with increasing $a / g$, though it is approximately zero for $a < g$ as expected since the acceleration is not strong enough for the particles to lift off the bottom plate.
Note that the particles can reach the top plate for $a / g \ge 1.2$.
The strength of the measured potentials has a non-monotonic and highly non-linear dependence on $\lambda$.
There is a threshold at $\lambda \sim 1$ with the plate potential being less than 20~mV for smaller $\lambda$.

\begin{figure}
	\subfloat[][]{\label{fig:cell_photograph} \includegraphics[width=8cm]{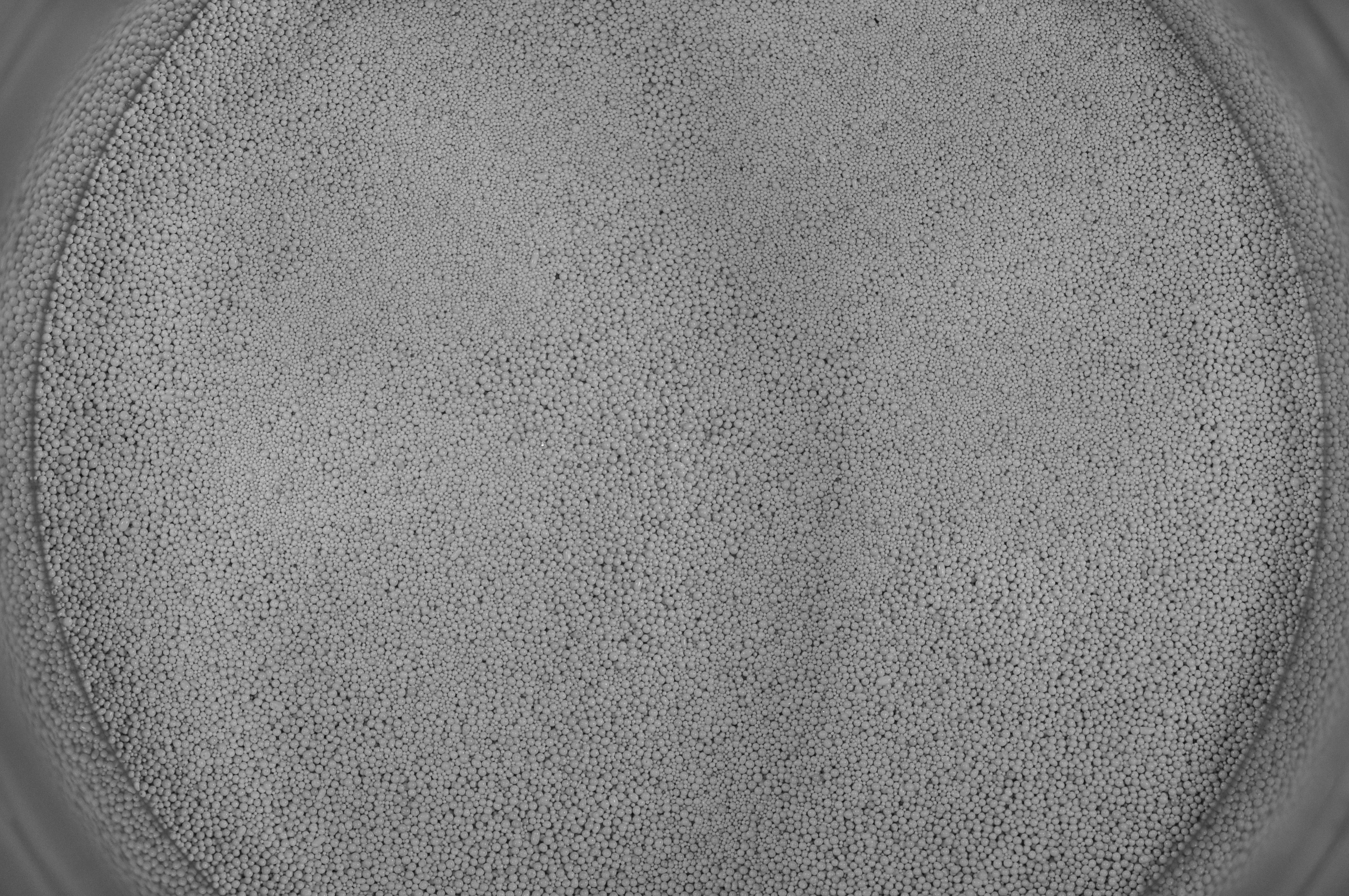}}
    \subfloat[][]{\label{fig:particle_mapping} \includegraphics{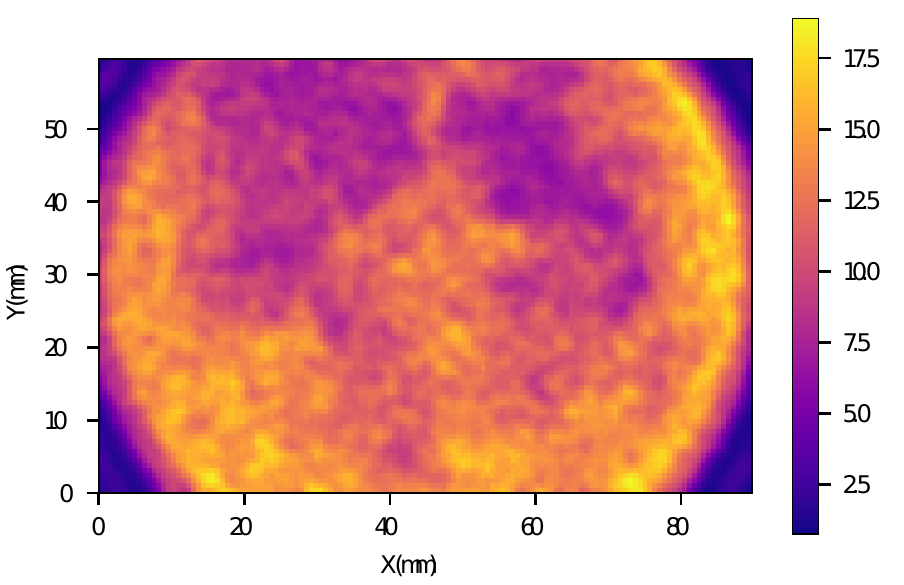}}
	\caption{Particle arrangement on the top layer after shaking. Top down view of the particles in the cell after being shaken for 20,000 cycles at $a = 2.08 \, g$. We used $\lambda = 3$ of 200--300~$\mu$m ZrO$_2$:SiO$_2$ and  $\lambda = 3$ of 400--600~$\mu$m ZrO$_2$:SiO$_2$ mixed together. \protect\subref{fig:cell_photograph} Photograph of the particles. \protect\subref{fig:particle_mapping} Map of which regions are predominately large particles (large values, bright patches) vs. small particles (small values, dark patches). The units for the values are arbitrary.}
	\label{fig:cell_particle_rearrangement}
\end{figure}

For the polystyrene spheres and glass spheres (Fig.~\ref{fig:lambda_dependence_three_materials}), the dependence on $\lambda$ also shows threshold behavior.
However, the dependence for these particles has a very different threshold of $\lambda \sim 10$ and the potential difference is $\approx 0.2$~V for $\lambda$ below the threshold.

\subsection{Evolving Particle Arrangement}

Collisional charge exchange effects due to dissimilar sized particles have already been documented.
Small particles tend to become negatively charged, while larger ones become positively charged \citep{Lacks_Sankaran_JPD_2011,Duff_Lacks_JournalElectrostatics_2008,Forward_etal_PRL_2009,Forward_etal_GRL_2009,Kok_Lacks_PRE_2009,Angus_etal_JournalElectrostatics_2013,Waitukaitis_etal_PRL_2014}.
That phenomenon, taken together with the well documented ability of granular flows to segregate by size when convecting, could lead to macroscopic charging \citep{Lacks_Sankaran_JPD_2011,Duff_Lacks_JournalElectrostatics_2008,Forward_etal_PRL_2009,Forward_etal_GRL_2009,Kok_Lacks_PRE_2009,Angus_etal_JournalElectrostatics_2013,Waitukaitis_etal_PRL_2014}.
Due to this feature of granular flows and the polydispersity of our particles, we investigated how particles of different sizes spatially segregate in the slab due to shaking.
To increase the potential segregation and to make it easier to identify, we mixed two size ranges of spheres of the same material together.

We mixed $\lambda = 3$ of 200--300~$\mu$m ZrO$_2$:SiO$_2$ and  $\lambda = 3$ of 400--600~$\mu$m ZrO$_2$:SiO$_2$ together and shook the cell for 20,000 cycles at $a = 2.08 \, g$.
Before and after the runs mixing two particle size ranges, the cell was disassembled and set on a flat surface carefully so as to not disturb the particle arrangement.
The particles were photographed (Fig.~\ref{fig:cell_photograph}) with a Nikon D90 camera and a Nikkor 110 mm macro lens from above at a resolution of $4288 \times 2848$~pixels, such that the full diameter of the cell would be completely visible ($D = 8.6$~cm corresponds to 4110 pixels)

A mapping is needed that will indicate thee regions where each particle size predominated.
By eye, regions dominated by larger particles can be distinguished from regions dominated by small particles by the larger and higher contrast gaps between the particles.
We used this to produce the map of the spatial pattern of which particle size dominates.

The photographed image was smoothed with a $7$~pixel radius disk to smooth over the contrast for regions dominated by small particles but not large particles.
Then, a map of the remaining contrast was found by dividing the image into $32 \times 32$~pixel squares and taking the standard deviation of the pixel values in each square.
Increasing the fraction of large particles in a region increases the value of this mapping.
The resulting map was then smoothed using a $2$~pixel radius disk to make the regions easier to see and not pick out single large particles in the middle of a region of small particles.

The mapping of which particle size dominates each area is shown in Figure~\ref{fig:particle_mapping}.
Particles of both sizes are visible on the top with patches that are dominated by each size and patches where they are mixed.
This indicates that the particles do not completely segregate and that there is complex macroscopic re-arrangement of the particles going on.

We repeated this procedure three more times, mixing the 200--300~$\mu$m ZrO$_2$:SiO$_2$ with each of the larger types of ZrO$_2$:SiO$_2$ spheres and found a similar pattern.
The same measurements were made with the different sizes of glass spheres, which showed similar patterns by visual inspection, but the same quantitative analysis could not be done reliably on the images we took.

\section{Conclusions}

We demonstrate a novel granular electrification experiment that mechanically shakes particles vertically in a cylindrical cell while measuring the macroscopic electric potential between the two endplates.
All 12 tested particle types exhibit electrification regardless of form (spheres or powder), size, conductivity and material, as seen by the non-zero electric potentials in Figs.~\ref{fig:cycle_profiles}.
Electric potentials for all particle types have similar shapes and amplitudes.
We found thresholds for electrification with respect to particle number (Fig.~\ref{fig:lambda_dependence}).
Additionally, when the polydispersity of the particles was increased by mixing two size ranges of the same material together, we found that the particles do not completely segregate and instead form rearranging patches of different mixing levels.
A macroscopic evolving charge pattern would be consistent with the long time-scale variations we measure in the electric potential, although challenging to measure.

In order to consider the implications of these results for granular electrification in general, we must assume the electrification and material exchange from plate-particle and wall-particle collisions is negligible compared to the electrification from particle-particle collisions.
The results then suggest that particle form, size, conductivity, and material are not fundamental to our observations of granular electrification.
Moreover, we argue that collective phenomena play a key role in the electrification.
That is not to say that two-body or multi-body collisions are irrelevant, but instead that the role of the large-scale structure and long-range interactions of large numbers of particles is pivotal.
A number of pieces of evidence point to this conclusion.
First, the threshold for electrification with increased particle number (Fig.~\ref{fig:lambda_dependence} and~\ref{fig:lambda_dependence_three_materials}) is suggestive of a phase transition.
Also, the long time dependence and inversions of the potential argue for complex macroscopic dynamics.
Collective phenomena being the dominant factor would predict relative independence of the electrification with respect to particle form, size, conductivity, and material like we observe.
Here we make analogy to the observation that gasses may exhibit similar thermodynamic behavior independently of their chemical composition (i.e. the ideal gas law is not material-specific).
One possible mechanism would be the spontaneous charging by local polarization recently found by \citet{Yoshimatsu_etal_ScientificReports_2017} assuming all of our particle types have a large enough polarizability and neutralization efficiency to show exponential growth in charging.

Now we consider the implications of collective phenomena on atmospheric electrification (thunderstorms, thunder-snow, volcanic lightning, and lightning in dust storms).
Our experiment is shaking a granular liquid and/or gas under a partial vacuum, which is a bit different than the more particle-dilute (except for volcanic lightning near the throat of the volcano) higher humidity turbulent granular-fluid flows found in the atmosphere.
Thunderstorms and volcanic ash clouds must be sufficiently tall to exhibit lightning  \cite{saunders_SpaceScienceReviews_2008,smirnov_PhysicsUspekhi_2014}.
Measurements of the electric fields inside thunderstorms show a complex spatial dependence over a large range of length scales \cite{marshal_etal_JGR_1995}.
Our observations that different materials all exhibit electrification parallel natural electrification in ash and dust.
Taken together, these suggest that collective phenomena are important in understanding natural storm electrification along side the effects of dissimilar size in charging \cite{Lacks_Sankaran_JPD_2011,Duff_Lacks_JournalElectrostatics_2008,Forward_etal_PRL_2009,Forward_etal_GRL_2009,Kok_Lacks_PRE_2009,Angus_etal_JournalElectrostatics_2013,Waitukaitis_etal_PRL_2014} and having more than one phase of the same material (e.g. H$_2$O) in the system \cite{saunders_SpaceScienceReviews_2008,smirnov_PhysicsUspekhi_2014}.

Further work is required to check the validity of the assumption that the electrification from plate-particle and wall-particle collisions is negligible compared to the electrification from particle-particle collisions.
Particles of different materials and/or size ranges need to be mixed together in the cell and their electric potential dynamics compared to the mono-material single size range case.
This will help better compare the effects of collective phenomena, material differences, and size differences on electrification.
Other avenues of future investigation would be using narrowing size ranges for the particles, finding out the cause for the different particle quantity thresholds (Fig.~\ref{fig:lambda_dependence_three_materials}), changing wall and plate materials, looking at the charge distributions of the particles themselves and/or the electric fields inside the cell, etc.

We argue that we can draw these conclusions about granular and atmospheric electrification without a detailed theoretical model of these phenomena, while that is surely desirable in the future.
The underlying difficulty is that of model complexity: these are far from equilibrium phenomena of two or more phase flows, often embedded in a background turbulent atmospheric flow.
This problem is conceptually more difficult than systems with known continuous equations of motion, such as fluid flow without particles.
Still, one can speculate on the nature of the theoretical possibilities that occur independently of material properties.
Finally, we hope that our experimental observations may help by conceptually framing atmospheric electrification as a type of macroscopic collective mechanism.
This highlights the need for  a better theoretical understanding of those processes.

\begin{acknowledgments}
	We acknowledge the past work on the project, including construction of much of the current apparatus, by and discussions with Paul Lathrop, Zack Lasner, Julia Salevan, Tyler Holland-Ashford, Eric Speiglan, and Allison Bradford.
	We thank our senior technicians Nolan Ballew and Don Martin for assistance in the construction and operation of the experiment.
	We thank John H. Abrahams III at the Center for Nanophysics and Advanced Materials at the University of Maryland College Park for the microscopy characterizing the powders.
	We thank Bruno Eckhardt, Seth J. Putterman, and Derek Richardson for discussions and comments, Daniel Serrano for discussions and editing, and Scott R. Waitukaitis for the suggestion to use 69\%:31\% ZrO$_2$:SiO$_2$ particles.
	This research was partially funded by the Julian Schwinger Foundation.
\end{acknowledgments}


\bibliography{granular_electricity,cloud_electrification,books,datasets}

%

\end{document}